\documentclass[default,iicol]{sn-jnl}

\usepackage{graphicx}%
\usepackage{multirow}%
\usepackage{amsmath,amssymb,amsfonts}%
\usepackage{amsthm}%
\usepackage{mathrsfs}%
\usepackage[title]{appendix}%
\usepackage{xcolor}%
\usepackage{textcomp}%
\usepackage{manyfoot}%
\usepackage{booktabs}%
\usepackage{algorithm}%
\usepackage{algorithmicx}%
\usepackage{algpseudocode}%
\usepackage{listings}%

\theoremstyle{thmstyleone}%
\theoremstyle{thmstyletwo}%

\theoremstyle{thmstylethree}%

\begin{document}

\title[The Gerasimov-Drell-Hearn sum rule with nuclear targets]{The Gerasimov-Drell-Hearn sum rule with nuclear targets}


\author*[1,2]{\fnm{Steven D.} \sur{Bass}}\email{Steven.Bass@cern.ch}

\author[3]{\fnm{Paolo} \sur{Pedroni}}\email{Paolo.Pedroni@pv.infn.it}

\author[4]{\fnm{Andreas} \sur{Thomas}}\email{thomand@uni-mainz.de}

\affil*[1]{
\orgname{Kitzbühel Centre for Physics}, \orgaddress{
\city{Kitzbühel}, 
\country{Austria}}}

\affil[2]{\orgdiv{Marian Smoluchowski Institute of Physics},
\orgname{Jagiellonian University}, \orgaddress{
\city{Krakow}, 
\country{Poland}}}

\affil[3]{
\orgname{INFN-Sezione di Pavia}, \orgaddress{
\city{Pavia}, 
\country{Italy}}}

\affil[4]{
\orgname{Institut f\"ur Kernphysik, University of Mainz, Mainz, Germany}}


\abstract{Hadron properties are modified when the hadron is embedded in a nuclear medium.
Here we discuss the Gerasimov-Drell-Hearn, GDH, sum rule for polarised photoproduction from a polarised nucleon within a polarised nuclear target. 
Strong enhancement is expected with the 
suppression of the proton and nucleon resonance masses and enhancement of the
proton's anomalous magnetic moment in medium. 
This could be tested in polarised photoproduction experiments
with interesting targets being polarised deuterons, $^3$He, $^6$Li and $^7$Li. 
The largest contribution to the GDH sum rule
comes from the $\Delta$ resonance excitation.
In existing data with polarised deuterons and $^3$He the $\Delta$ 
excitation is shifted to slightly lower energy when compared to model predictions where the $\Delta$ is treated with its free mass.}




\maketitle

\section{Introduction}

Hadron properties are modified in medium with partial restoration of chiral symmetry 
\cite{ericson,Metag:2017yuh,Kienle:2004hq,Bass:2018xmz,Saito:2005rv}.
Hadron masses, the pion decay constant,
and the nucleon's axial charge
and magnetic structure behave as dependent on the nuclear medium.
In high energy deep inelastic scattering
the EMC nuclear effect 
tells us that the QCD parton structure of the proton is modified when the proton is in a nucleus \cite{Cloet:2019mql}.
Via the Bjorken sum rule,
quenching of the nucleon's axial charge
$g_A^{(3)}$ 
in medium, $\approx 20\%$ in large nuclei
\cite{ericson,Ericson:1998hq,Suhonen:2018ykq},
means that the nucleon's internal spin structure 
\cite{Bass:2004xa,Aidala:2012mv}
will also be modified in medium;
for recent discussion see, e.g., 
\cite{Bass:2020bkl,Thomas:2018kcx}.

In this paper we focus on medium dependence of the nucleon's spin structure measured through polarised photoproduction.
One expects nuclear  dependence of the spin-dependent photoabsorption 
cross sections for photons scattering on bound nucleons. 
We make the first observation of a nuclear medium effect in 
present data with deuteron and $^3$He targets.
In the spin cross sections where the photon and target are polarised parallel
one sees that the 
$\Delta$ resonance peak is shifted to lower energies by up to $\approx -20$ MeV. 
The effect is qualitatively different to that expected from smearing due to Fermi motion of the bound nucleons and not seen in the spin averaged cross section. 
More generally, one also expects modification of the value of the 
Gerasimov-Drell-Hearn, GDH, sum rule 
for bound nucleons with circularly polarised real photon beams scattering from longitudinally polarised nuclear targets.

The GDH sum rule for polarised photoproduction
reads~\cite{Gerasimov:1965et,Drell:1966jv}:
\begin{equation}
\int^{\infty}_{M^2} 
\frac{ds_{\gamma A}}{s_{\gamma A} - M^2}
(\sigma_{\rm p} - \sigma_{\rm a})
= 4 S \pi^2 \alpha_{\rm QED} \kappa^2 / M^2 \ ,
\end{equation}
where $\sigma_{\rm p}$ and $\sigma_{\rm a}$ 
are the spin-dependent photoabsorption cross sections
involving photons polarised parallel and antiparallel to the target's spin.
Here
$s_{\gamma A}$ 
is the photon-target centre-of-mass energy squared
with $\kappa$ the target's anomalous magnetic moment;
$M$ is the target mass and $S$ is its spin.
The GDH sum rule is derived from the very general principles of causality, unitarity,
Lorentz and electromagnetic gauge invariance
together with the single assumption that 
$\sigma_{\rm p} - \sigma_{\rm a}$
satisfies an unsubtracted dispersion relation.

For free protons with $\kappa=1.79$ 
the GDH sum rule predicts a value of $205 \ \mu$b
for the integral in Eq.~(1).
For neutrons with anomalous magnetic moment  -1.91 
the sum-rule predicts a GDH integral of $232 \ \mu$b.

When considering meson photoproduction on
bound nuclei both sides of the GDH sum rule 
are expected to be enhanced in medium~\cite{Bass:2004xa}.  
The sum rule involves the target anomalous magnetic moment squared divided by the target mass squared. 
The nucleon and 
P$_{33}$ $\Delta (1232)$ resonance masses 
are suppressed in medium 
\cite{ericson,Saito:2005rv,Oset:1987re,GarciaRecio:1989xa,Mosel:2020zdw}.
Nucleon magnetic moments are expected to be enhanced 
\cite{Lu:1998tn}
with possible evidence observed in JLab data  \cite{Plaster:2005cx}. 
Thus, medium dependence occurs four times on the right hand side of Eq.~(1) making the GDH sum rule
especially sensitive to possible medium effects. 
Reduction in the $\Delta$ mass means that the $\Delta$ excitation
contribution to the spin cross section difference 
$\sigma_{\rm p}-\sigma_{\rm a}$ will occur at slightly lower incident photon energies.
One expects a factor of up to about two enhancement in the GDH integral at nuclear matter density with sizeable effects that might be looked for also in finite nuclei in ongoing and future experiments.

The plan of this paper is as follows.
Section 2 describes present measurements of the proton GDH sum rule. 
In Section 3 we describe the theoretical expectations for how the GDH sum rule might be modified in medium, e.g., 
with polarised
photoproduction from a polarised nucleon in a polarised nucleus. Here careful control is needed in realistic experiments
relating the spin of the struck nucleon to the polarisation of the nucleus. 
In Section 4 
we discuss the situation with polarised deuteron, $^3$He, $^6$Li and $^7$Li 
as well as with possible heavier targets 
for use in possible future experiments, e.g., at Mainz, Bonn and JLab. 
In Section 5 we point to present data which suggest 
a possible shift of the $\Delta$ excitation contribution to the GDH integral to lower incident photon energies.
Finally, in Section 6 we make our conclusions.

\section{The GDH sum rule for the proton}

So far GDH experiments have been done in fixed target mode, so it is common to work in the laboratory frame,
$s_{\gamma p} = (q+l)^2 = 2M \nu +M^2$,
with $\nu$ the photon 
energy, $M$ the target proton mass and $l$ and $q$ the target and photon four-momenta.

Experiments at Bonn (ELSA) and Mainz (MAMI) have measured
the GDH integrand 
for a proton target
through the range of incident photon energies 
$\nu =$ 
0.2 - 0.8 GeV (MAMI) 
and 0.7 - 2.9 GeV (ELSA) \cite{Ahrens:2001qt,Dutz:2003mm,Dutz:2004zz}.
The inclusive cross-section for the proton target
${\sigma_{\rm p} - \sigma_{\rm a}}$
is dominated by the $\Delta$ resonance 
excitation 
with contribution $\sim +190 \ \mu$b to the GDH integral from photon energies between 200 and 400 MeV.
This is primarily the spin-flip M1
($M_{1+}$)
magnetic transition 
which dominates over the 
electric quadrupole E2 
($E_{1+}$)
amplitude  
\footnote{In the multipole notation inside the parentheses,
    the first subscript denotes the
    the orbital angular momentum $l_{\pi}$ of the photoproduced pion
    in the $\gamma N\to \Delta\to N\pi$ reaction
    and  the sign $\pm$  refers to the two possibilities to construct the total $N\pi$
    angular momentum $J= |l_{\pi} \pm 1/2|$.}.
One also observes smaller resonance contributions from the 
D$_{13} (1520)$ and F$_{15} (1680)$ and
F$_{35} (1905)$
nucleon excitations 
\cite{Helbing:2006zp,Drechsel:2004ki,Pedroni:2011zz}.

The corresponding integral from 
200 MeV up to 2.9 GeV,
$\sqrt{s_{\gamma p}} = 2.5$ GeV,
was thus extracted from proton fixed target experiments.
One finds
\cite{Dutz:2004zz,Pedroni:2011zz}:
\begin{equation}
  \int^{2.9 \ {\rm GeV}}_{0.2 \ {\rm GeV}} 
\frac{d \nu}{\nu}
(\sigma_{\rm p} - \sigma_{\rm a})
= +253.5 \pm 5 \pm 12 
 \ \mu {\rm b} \ , 
\end{equation}
with $\nu$ the incident photon energy in the laboratory frame.
The contribution to the 
sum rule from the unmeasured region close to threshold
between 140 and 200 MeV is estimated from the 
MAID~\cite{
Drechsel:2004ki,Kashev},
SAID~\cite{A2:2019yud}
and 
BNGA~\cite{Anisovich:2017afs}
models 
as
$-29.5 \pm 2$ $\mu$b 
when we average over the latest versions of the different model predictions.

For the higher energy part,
one presently uses estimates deduced from high energy low $Q^2$
data with the most precise measurements coming from 
CLAS \cite{Fersch:2017qrq}
and COMPASS \cite{Aghasyan:2017vck}
with photon-proton centre of mass energies between 2.5-2.9 and 11-15 GeV respectively.
No $Q^2$ dependence is visible in finite, 
e.g. non-vanishing, 
proton asymmetries below 
$Q^2=0.5$ GeV$^2$. 
From a Regge motivated fit to these low $Q^2$ data one estimates the
high-energy contribution to the GDH sum rule from 
$\sqrt{s}_{\gamma p} \geq 2.5$ GeV to be 
\cite{Bass:2018uon}:
\begin{equation}
\int^{\infty}_{2.9 \ {\rm GeV}} \frac{d \nu}{\nu} 
(\sigma_{\rm p} - \sigma_{\rm a}) 
= - 15 \pm 2 \ \mu {\rm b} .
\end{equation}
High energy deuteron asymmetries are consistent with zero
in the low $Q^2$ data within the same kinematics
\cite{CLAS:2015otq,Compass:2007qxf}
meaning that the high-energy contribution is coming predominantly from the isovector channel.

Combining Eq.~(3) with the integral contributions from threshold up to 
photon energies of 2.9 GeV gives 
\begin{equation}
\int^{\infty}_{{\rm threshold}} 
\frac{d \nu}{\nu}
(\sigma_{\rm p} - \sigma_{\rm a})
= + 209 \pm 13 \ \mu {\rm b} 
\end{equation}
for the proton GDH sum rule.

CLAS have made an independent check of the GDH sum rule by extrapolating
inclusive data at 
low $Q^2$, between 0.012 and 1 GeV$^2$, to the photon point.
They obtain the result 
$+204 \pm 11 \mu$b 
\cite{CLAS:2021apd}.
 GDH experiments have also been performed by the LSC collaboration
with a main focus on inclusive 
photoproduction of 
$\pi^0$ 
with $\nu$ between 200 and 420 MeV~\cite{LSC:2008wiu}.

\section{The GDH sum rule in medium - theoretical considerations}

Both sides of the GDH 
sum rule are expected to be enhanced
in medium above pion production threshold. 
The nucleon and resonance, 
including the $\Delta$, 
masses and the nucleon magnetic 
moments are expected to change in nuclei.

When discussing the GDH sum rule in medium one implicitly assumes the validity of the dispersion relation for bound nucleons not on their mass shell. 
The same assumption is 
made in almost all discussions of the EMC effect for deep inelastic scattering from nuclear targets, with the parton model built on the dispersion relation for forward Compton
scattering, the light-cone operator product expansion and QCD factorisation, 
see, e.g., Ref~\cite{Bass:2004xa}. 
\footnote{The GDH sum rule has been 
explicitly shown to work for a virtual polarised photon target 
with fixed virtuality to all orders in perturbation theory \cite{Bass:1998bw}.}

Chiral models give a $\Delta$ 
mass shift in medium of 
$-33 \times \rho/\rho_0$ MeV 
with $\rho$ the nuclear density and $\rho_0$ the density of nuclear matter.
In the same models 
the proton mass shift is about
1.5 times bigger with range expected in -50 to -40 MeV
\cite{ericson,Oset:1987re,GarciaRecio:1989xa,Mosel:2020zdw}.
This nucleon-$\Delta$ mass shift is driven predominantly by 
the colour hyperfine interaction 
or one gluon exchange potential \cite{Close-book}.

If one takes $\sim -45$ MeV as a good estimate of the nucleon mass 
change in medium at $\rho_0$, then 
one picks up an enhancement factor of
1.1 on the right hand side of the GDH integral, Eq.~(1),
from the mass denominator.
For the anomalous magnetic moment we 
make a first estimate relating medium changes in magnetic structure to the nucleon's axial charge with 20\% quenching in $g_A^{(3)}$ at nuclear matter density.
For the magnetic moment 
we assume the leading-order (before pion cloud effects) relation obtained in the quark meson coupling model
\cite{Saito:1994kg}, 
\begin{equation}
\mu^*_N / \mu_N 
 \sim
g_A^{(3)}/g_A^{* (3)} \ ,
\end{equation}
where the superscript * denotes the in medium quantity and 
with caveat 
that the model gives just $\approx 10\%$
suppression of $g_A^{* (3)}$
at nuclear matter density. 
The key physics input here is that 
constituent quark mass 
decreases
in medium so quarks behave as more relativistic and the lower P-wave component of the quark wavefunction is enhanced.
Then with 20\% quenching of $g_A^{(3)}$ at nuclear matter density
one gets a 
1.7 factor enhancement 
from the proton's anomalous magnetic moment squared, 
so giving a net enhancement of about 1.9 in the GDH integral.
\footnote{
Here we have taken the change in magnetic properties as dependent on the nuclear density.
With the less bound valence nucleons carrying the polarisation, one might wonder whether this affects theoretical predictions. 
For deep inelastic scattering 
smaller effects for valence nucleons were suggested in the EMC effect model 
of \cite{CiofidegliAtti:1999kp}
whereas shadowing effects at small Bjorken $x$
were found to be 
enhanced when a single nucleon carries the measured quantity  \cite{Guzey:1999rq}.
}

Given that the GDH sum rule is working for bound nucleons, 
medium modifications should also be manifest in the 
spin cross section 
$\sigma_{\rm p}-\sigma_{\rm a}$.
The $\Delta$ excitation contribution to the GDH integral 
should be shifted to smaller $\Delta$ 
excitation energy, 
weighted by $1/\nu$ in the integral.
Smaller, higher-mass,  resonance contributions to the sum-rule will, in general, also be subject to mass shifts.
The $\Delta$ width will be enhanced with the 
area under $\sigma_{\rm p}-\sigma_{\rm a}$
spread out in energy. 
Above the resonance mass, 
the weight shifted to smaller energies will have a bigger effect in the GDH integral whereas the
weight shifted to higher energies will have reduced effect in the integral.
For the idealised case of Breit-Wigner
these shifts are approximately symmetric in
$\sigma_{\rm p}-\sigma_{\rm a}$
about the resonance mass
but effects at
smaller energies,
closest to the $1/\nu$ ``pole'' in the GDH integrand, 
Eq.~(2), 
will have the bigger effect in the integral.
There will also be small contributions
from change in pion production threshold with change in the nucleon mass $M$.
The experimental challenge is to measure these contributions.
We explain in Section 5 
what is so far seen in present data with deuteron and $^3$He targets.

The Regge intercepts for high-energy
polarised photoproduction are not expected to be target dependent and are properties of the exchanges rather than the targets. 
Regge intercepts describe the asymptotic high energy behaviour of scattering amplitudes.
For unpolarised scattering
target independence is illustrated in \cite{Landshoff:1994up} 
where one observes the same intercepts describing 
photoproduction, pion-proton and proton-antiproton collisions.

\section{Choice of target}

Spin polarised targets have been investigated since decades to get the best figure of merit for particle physics experiments \cite{Goertz:2002,Thomas:2006apt,Thomas:2011apt,Dutz:2017}.  
Technically spin polarised targets are realised as cryogenic solid-state targets using the dynamical nuclear polarisation (DNP) technique \cite{Abragam:1962},  by the ``brute force'' method in the ``HD-ice'' target \cite{LSC:2008wiu}, 
or the use of optical pumping for gas targets. 

Besides the degree of polarisation, 
a very important parameter for a solid spin polarised target is the dilution factor, which is the ratio of polarisable nucleons to the background nucleons. To provide highly spin polarised solid targets, chemical compounds are most often used instead of pure elements. This led to the use of Butanol (${\rm C_4H_9OH}$, dilution factor 10/74 = 0.135) or Ammonia (${\rm NH_3}$, dilution factor 3/17 = 0.176) as proton targets. 
Small pieces of target material have to be filled into a target container and cooled by a cryogenic liquid (typically 50\% of $^4$He and/or $^3$He), leading together with technical windows of the refrigerator to an additional dilution.

Natural choices as ‘quasi-neutron’ target are their deuterated equivalents. As an alternative a $^3$He gas target was used at MAMI for the measurement of the GDH observable on the neutron \cite{Krimmer:2011zz,AguarBartolome:2013mga}. 
The low density was an important boundary condition, since the maximum photon flux was limited by the photon energy tagging system and the target length had to match the detector acceptance. On the other hand, the ratio of events produced on the $^3$He gas to the target cell windows had to be optimised, in other words the window thickness had to be minimised for a target pressure optimal for highest spin polarisation.

The need for a better dilution factor and radiation resistance led to the development of Lithium compounds as target material, first in Saclay \cite{Ball:2004} and later for the CERN COMPASS \cite{Goertz:1995} and the SLAC E155 \cite{Bueltmann:1999} experiments.
 At CERN $^6$LiD was used for the experiment \cite{Ball:2003}, $^7$LiH was investigated in parallel in the laboratory \cite{Meier:2001} to learn about the behaviour of the different spin species in the compound. 
This gives the possibility to use these light nuclei ($^6$Li and $^7$Li) as target material for the investigation of the GDH observable in photon induced reactions.

One might consider also heavier nuclear targets.
Many nuclei with non-zero spin can be polarised using the DNP technique. Additionally, optical pumping has been used to polarise heavy nuclei, notably Xenon.
For heavier nuclei the 
polarisation is approximately carried by a valence nucleon
leaving the net larger 
part of the nucleus spin independent.
This leads to an additional $1/A$ spin dilution factor with $A$ the atomic number.
Experimentally, 
some compromise between  the nuclear density and this dilution factor has to be made.
For the JLab polarised deep inelastic experiment on nuclear targets, this compromise converged on $^7$Li \cite{JLab-Li}.

Table \ref{gdhparam}
summarises the more relevant properties of spin-one deuteron,
$J=\frac{1}{2}$ $^3$He 
and larger spin $J=1$ $^6$Li and
$J=\frac{3}{2}$ $^7$Li 
target nuclei as well as 
$^{129}$Xe as a representative heavy nucleus, 
giving their magnetic and anomalous magnetic moments and corresponding predictions for their GDH integrals. 
The anomalous magnetic moment $\kappa$ for each target is related to the  magnetic moment $\vec{\mu}$ through 
$
\vec{\mu}
= \frac{e}{M} (Q+\kappa) \vec{S}$, 
where $Q$, $M$ and $\vec{S}$ are the target charge, mass and spin.
Table \ref{gdhparam} also gives, for the same nuclei, 
{
the estimated effective degree of spin polarisation  (i.e., the expectation value for the $z$ component of the spin, summed over all particles of the same type) for both protons, $P_p$, and neutrons, $P_n$, 
which are reduced with respect to the free case due to the nuclear structure \cite{Wiringa:2013ala, Dzuba:2007zz}. }
These spin polarisations are important in
the selection of nuclear targets and
extraction of the key observables.

For nuclei, the GDH sum rule receives contributions from both nuclear photo-disintegration processes  and scattering 
from bound nucleons in medium.
One needs to subtract off the photo-disintegration part to determine the part of the sum rule from scattering on bound nucleons, 
with energies above the pion production threshold sensitive to the bound proton and neutron structure. 
Medium effects on individual bound nucleons will be larger the bigger the target nucleus.

\begin{table}[t!]
%
\caption{
GDH parameters for free nucleons and
different target nuclei. 
The magnetic moments $\mu$ are given
in units of nuclear magnetons, 
$\kappa$ are the anomalous magnetic moments 
and the GDH integrals 
$I_{\rm GDH}$ are quoted in units of $\mu$b.
The estimated effective nucleon polarisations
$P_{p}$ and $P_{n}$ for protons and 
neutrons are quoted for 
the different target nuclei. 
Values for the deuteron (see Eq. \ref{eq:deut}) are 
evaluated assuming a D-state probability of 5\%;
$^3$He and $^{6,7}$Li 
are taken from~\protect\cite{Wiringa:2013ala} 
and $^{129}$Xe from
\cite{Dzuba:2007zz}.
\label{gdhparam}}
\begin{tabular}[t]{c|lllll}
\hline
& $\mu$ & $\kappa$ & $I_{\rm GDH}$ & $P_p$ & $P_n$
\\
& & & ($\mu$b) & & \\
\hline
$p$  & +2.79 & +1.79 & 205 \\
$n$ & -1.91 & -1.91 &
232 \\
$d$   
& +0.86 &  -0.14 & 0.65 & 0.925 & 0.925 \\
$^3$He 
& -2.13 & -8.37 & 498 & -0.052 & 0.876 \\
$^6$Li & +0.82 & -0.55 & 1.08
& 0.848 & 0.848 \\
$^7$Li  
& +3.26 & +4.57 & 82.2
&
0.868 & -0.038 \\
$^{129}$Xe 
& -0.78 & -153.5 & 91.5 & 0.24 & 0.76 \\
\hline
\end{tabular}
%
\end{table}

\subsection{Deuterons}

The deuteron anomalous magnetic moment in the GDH sum rule corresponds to both  photo-disintegration of the deuteron as well as 
contributions from scattering on the bound proton and neutron.
The latter part of GDH integral describing proton and neutron structure is given as
\begin{equation}\label{eq:deut}
 {\rm GDH}^p +{\rm GDH}^n \approx
 {\rm GDH}^d / (1-1.5 \omega_D)  \ ,
\end{equation}
where $\omega_D = 5 \pm 1$\% is a small D-state probability in the deuteron
\cite{Machleidt:1987hj}.

\subsection{$^3$He}

$^3$He behaves, at first order, like a spin zero combination of two protons plus a spin half neutron carrying the spin of the nucleus 
with effective
nucleon polarisations
$P_n = 0.876$ 
and
$P_p = - 0.052$ 
determined by the combination of the predominant S-state (90\% probability) with the lower probability S$\prime$ and D states 
with numbers quoted from variational 
Monte-Carlo calculations in
\cite{Wiringa:2013ala}.

\subsection{$^6$Li and $^7$Li}

$^6$Li behaves like a spin-zero $\alpha$ particle 
(with two protons and two neutrons in the 1s shell)
plus a proton and neutron carrying the spin one of the nucleus.
These valence nucleons can also be in a P-state unlike for the deuteron.
For $^6$Li one finds 
$P_p=P_n=0.848$ 
for the polarisation of both protons and neutrons, so each carrying $\approx 42\%$ of the spin one of the nucleus
\cite{Wiringa:2013ala}.

For $^7$Li one has two protons and two neutrons in the 
1s shell with total spin zero 
with the other nucleons in the 1P$_{3/2}$ shell. 
In standard shell model calculations \cite{Landau:1991wop}
one finds $P_p=13/15$ and $P_n=2/15$.
More precise microscopic model calculations give
$P_p =0.868$
and
$P_n= -0.038$
\cite{Wiringa:2013ala}. 
With most of the nucleus' polarisation 
carried by the valence proton, 
$^7$Li is most suitable for a direct comparison with free proton data, whereas $^6$Li compares more directly with the isoscalar deuteron.

In nuclei the $\Delta$ width is significantly larger than in the free case and also the shape is distorted by final state interactions and other nuclear effects. 
{Given the energy resolution already achieved with the existing photon tagging systems ($\sim 2$ MeV, that can be lowered well below one MeV over a reduced energy range, see \cite{reiter:2006}), a 10-20 MeV peak shift can}
be detected in future experiments, similar to the situation observed with deuterium and $^3$He targets discussed below.

As the larger nucleus, 
Lithium is closer to the domain of mean field model applications.
For polarised deep inelastic scattering at large $Q^2$ a spin EMC effect is predicted for $^7$Li targets in 
\cite{Cloet:2006bq,deBarbaro:1984gh,Guzey:1999rq}
with
$g_A^{(3)}$ 
quenching of about 36\% the nuclear matter density prediction found in a confining Nambu-Jona-Lasinio, NJL, model calculation
\cite{Cloet:2006bq}.

\section{Existing data with deuterons and $^3$He}

\begin{figure}[t]
\centering
\includegraphics[width=0.52\textwidth]{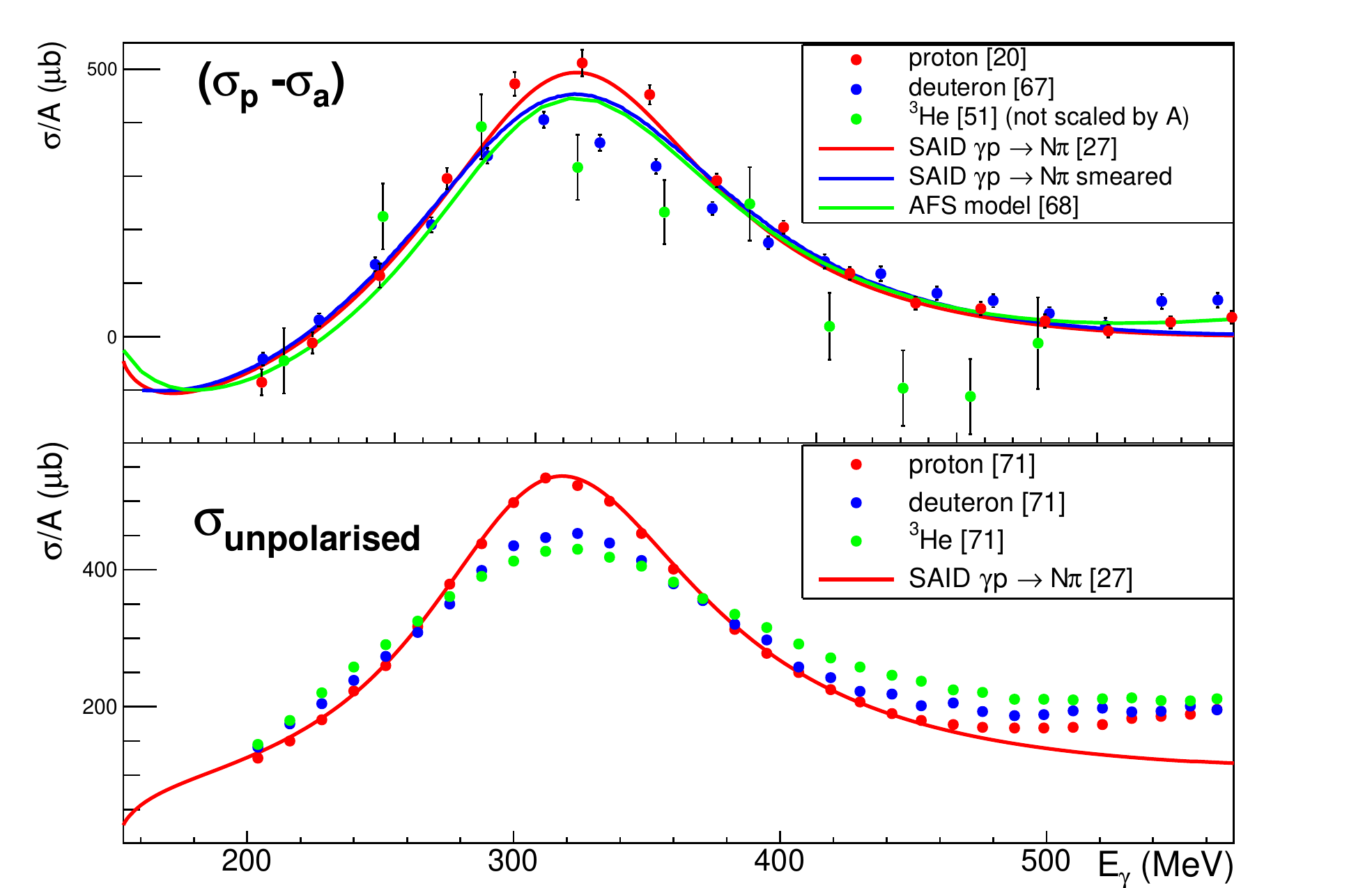}
\caption{
{Upper plot: The measured measured total inclusive helicity dependent cross sections for proton \cite{Ahrens:2001qt} , deuteron \cite{Ahrens:2009zz} and $^3$He \cite{AguarBartolome:2013mga} in the $\Delta$ resonance region are compared to the predictions of the deuteron AFS model \cite{Arenhovel:2004ha} and to the results of the SAID $N\pi$ analysis \cite{A2:2019yud} both for free proton and when the smearing due to deuteron Fermi motion is applied.
Lower plot: The unpolarised total inclusive cross sections for proton, deuteron and $^3$He \cite{MacCormick:1996jz} are shown together with 
the results of the SAID $N\pi$ analysis \cite{A2:2019yud} for the free proton case.
}
}
\end{figure}

First photoproduction 
measurements on polarised nuclear targets have been carried out using deuterons and $^3$He.
Incident photon energies ranged from 200 MeV up to about 1800 MeV with the deuteron
\cite{Ahrens:2009zz}
and up to 500 MeV with $^3$He \cite{AguarBartolome:2013mga}.
\footnote{
We note also TUNL measurements up to 29 MeV 
below the pion production threshold with $^3$He \cite{Laskaris:2020ddr}.
}

First Mainz data for the deuteron gave \cite{Ahrens:2009zz}:
\begin{equation}
  \int^{1.8 \ {\rm GeV}}_{0.2 \ {\rm GeV}} 
\frac{d \nu}{\nu}
(\sigma_{\rm p} - \sigma_{\rm a})
= +452 \pm 9 \pm 24
 \ \mu {\rm b} \ , 
\end{equation}
with tendency of a  still slowly rising integral with increasing 
upper energy at the measured limit of 1.8 GeV.
This compares with the theoretical prediction for the 
sum of proton and neutron GDH integrals in free space
$
\int^{\infty}_{\rm threshold} 
\frac{d \nu}{\nu}
(\sigma_{\rm p} - \sigma_{\rm a})
= +437 
 \ \mu {\rm b} 
$. 
Taking into account the D-state $\omega_D$ factor gives an expected
contribution to the deuteron GDH integral 
from scattering on bound nucleons of 404 $\mu$b.
Close to threshold contributions are discussed in
\cite{Arenhovel:2004ha}
involving extra channels compared to a proton target  
and cancellation between
photo-disintegration and meson production contributions.
From high energies one expects just a very small
contribution based on the 
low $Q^2$ asymmetries measured by CLAS \cite{CLAS:2015otq}
and COMPASS \cite{Compass:2007qxf}.

Interestingly, 
in the published deuteron data 
\cite{Ahrens:2009zz}  
the $\Delta$ excitation contribution to 
$\sigma_{\rm p}-\sigma_{\rm a}$
appears to be shifted 
by incident photon energy up to $\approx -20$ MeV 
compared to theoretical model predictions 
\cite{Arenhovel:2004ha} 
where the proton and $\Delta$ masses are taken with their values in free space. 
This effect is shown in the upper plot of Fig.~1, 
which displays the 
helicity dependent total cross sections for proton, deuteron and $^3$He targets. 
One observes that the Arenh\"ovel et al model 
(AFS) 
\cite{Arenhovel:2004ha}
in the $\Delta$ region basically follows the
  ``simple'' smearing just due to Fermi motion but the data show a different
  behaviour.
Even if statistics are poorer, with 
 $^3$He one also observes a small peak shift, $\approx -20$ MeV to lower incident photon energies
\cite{AguarBartolome:2013mga}.
No clear peak shift is observed in the 
spin averaged cross section  $(\sigma_{\rm p}+\sigma_{\rm a})/2$ for both 
deuteron and $^3$He, as  shown in the lower plot of Fig.~1.

\begin{figure}[t]
\centering
\includegraphics[width=0.52\textwidth]{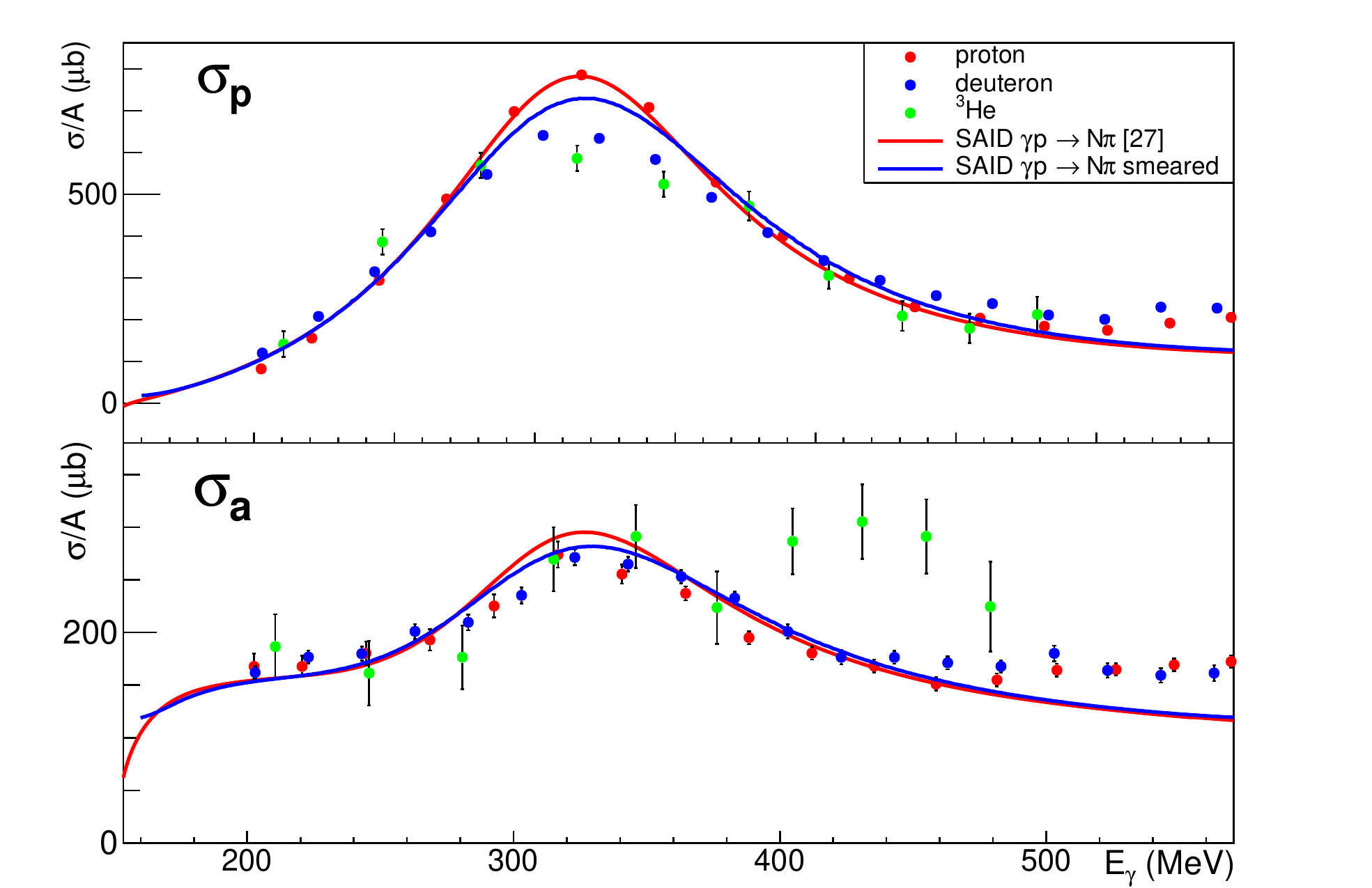}
\caption{ 
{
The total inclusive cross sections  $\sigma_{\rm p}$ and $\sigma_{\rm a}$
for proton, deuteron and $^3$He  are compared to
the results of the SAID $N\pi$ analysis \cite{A2:2019yud} both for free protons and when the smearing due to deuteron Fermi motion is applied.
}
}
\end{figure}

Fig.~2 shows the spin-dependent cross sections
$\sigma_{\rm p}$ and $\sigma_{\rm a}$ 
for the proton, deuteron and $^3$He, as  determined by combining the two previous observables. 
It can be noticed that the peak energy is shifted downwards 
in the parallel spin cross section
$\sigma_{\rm p}$ 
associated with the $\Delta$ excitation 
in both the deuteron and $^3$He data, while  
with no peak shift in the antiparallel spin cross section $\sigma_{\rm a}$ is visible.

For heavy nuclei,
the peak in the spin averaged cross section is shifted to higher energy
with damping and increased width - see Fig.~3 
which shows the unpolarised total inclusive cross sections for proton, $^{12}$C and $^{208}$Pb targets.
This feature is explained by the so-called $\Delta$-hole model 
\cite{Oset:1981wj,ericson}, 
i.e. by the propagation
of the $\Delta$ resonance
inside the nuclear matter, with effect much more important
than the one due to Fermi motion.

The downwards shift in the $\Delta$ 
peak energy observed in $\sigma_{\rm p}$
is a different effect to the Fermi smearing and $\Delta$ propagation in the nucleus. 
These observations suggest a challenge for new investigation: 
will the mass shift of the
$\Delta$ excitation contribution
survive more accurate data and might we see a more enhanced contribution to the GDH sum-rule with a larger polarised nuclear target  like $^6$Li or $^7$Li?
For the deuteron case 
binding effects, about 2 MeV, are smaller than the   observed shift. Moreover, they play the same role in both the spin antiparallel and parallel cross sections.

Small, few percent, medium modifications of
nucleon properties in light deuterons are also
observed in experimental measurements of the
EMC nuclear effect where parton distributions
of bound nucleons in the deuteron are seen in
experiment to be modified relative to free protons 
\cite{Griffioen:2015hxa}.
Also, 
while the deuteron is too small and difuse for application of mean field models, 
small changes in the values of the nucleon's axial and tensor charges 
in light nuclei including the deuteron are reported in recent lattice calculations \cite{Chang:2017eiq}.
A small shift in the $\Delta$ excitation is also seen in low energy processes 
involving proton-proton, proton-deuteron and pion-deuteron reactions,
see e.g., \cite{Hoshizaki:1992uh,Oh:1997eq}.

JLab low $Q^2$ measurements on the deuteron \cite{CLAS:2017ozc} and $^3$He 
\cite{JeffersonLabE97-110:2019fsc} when extrapolated to the photon point
give a neutron GDH integral consistent with theory 
to 20\% or 1$\sigma$  accuracy.
This uncertainty certainly includes the possible medium dependence discussed here within the experimental errors.

The GDH sum rule with nuclear targets is probing  change in the nucleon's magnetic structure in medium. 
If we believe we understand the nucleon mass shift from, e.g.,
chiral models, then any change in the static right hand side
of the sum-rule would point to modifications of magnetic
structure via the anomalous magnetic moment.
Change in $\Delta$ resonance excitation 
contribution to the $\sigma_{\rm p}$ spin cross section corresponds to a change in the magnetic structure through the M1 magnetic transition.

\begin{figure}[t]
\centering
\includegraphics[width=0.52\textwidth]{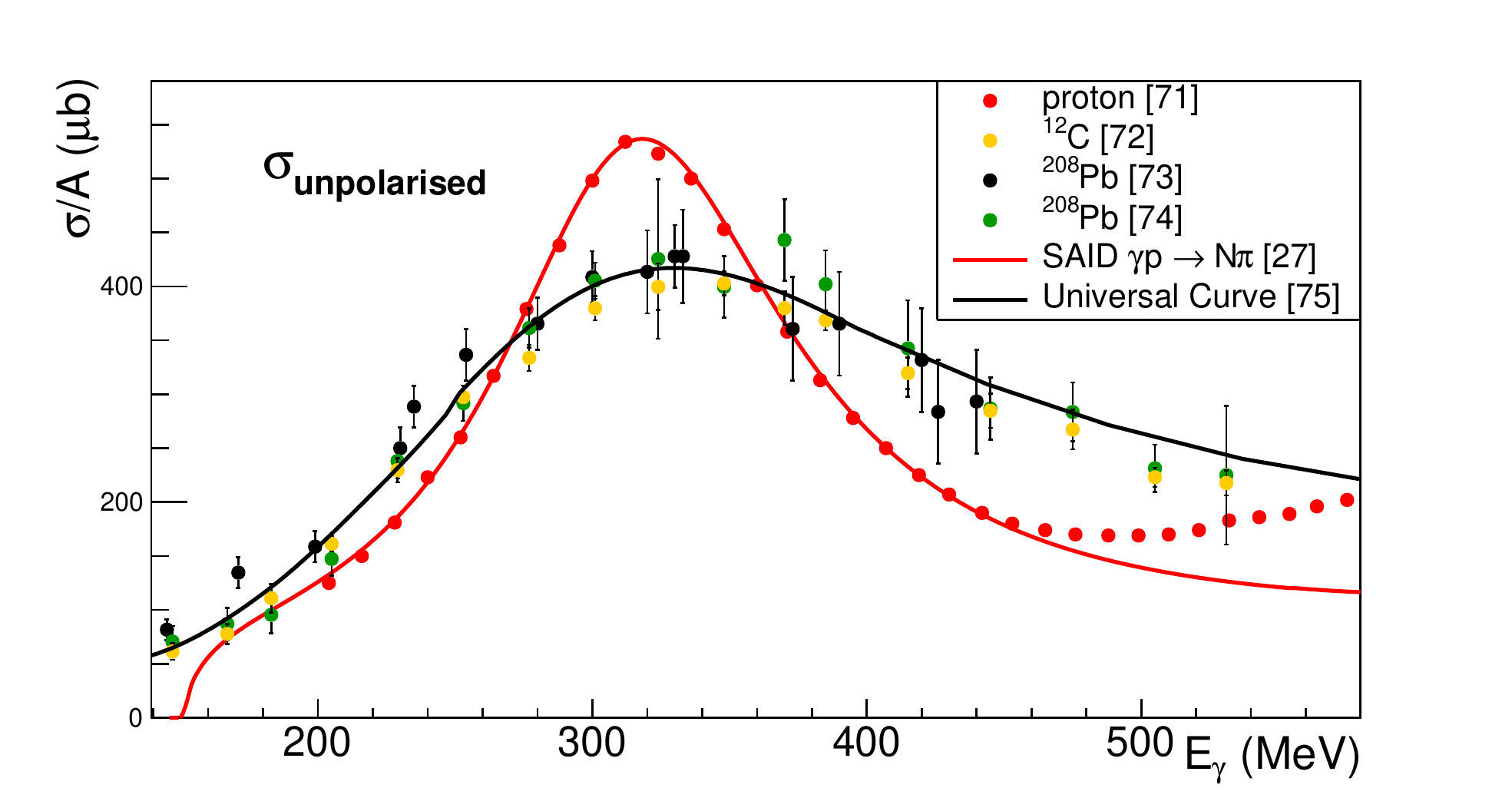}
\caption{ 
{
The unpolarised total inclusive cross sections for proton \cite{MacCormick:1996jz}, 
$^{12}$C \cite{Mecking:1979xs} and $^{208}$Pb
\cite{Carlos:1984lvc, Ghedira:1984}
are shown together with 
the results of the SAID $N\pi$ analysis \cite{A2:2019yud} for the free proton case and the co-called ``universal curve'', an average behavior of several published results for medium and heavy nuclei \cite{MacCormick:1997ek}.
}
}
\end{figure}

\section{Conclusions}

The GDH sum-rule for polarised nucleons embedded in polarised nuclei 
might be enhanced by a factor up to about two at nuclear matter density, with sizable effects also waiting to be investigated in lighter nuclei from the deuteron up to $^7$Li.
The effect could be up to an order of magnitude larger than the quenching of 
$g_A^{(3)}$
relevant to the first moment of the proton's deep inelastic structure function measured at large $Q^2$.

There are hints in first deuteron and $^3$He target data from Mainz of a shift 
in the $\Delta$ excitation contribution 
to lower incident photon energies.
This effect in the parallel spin cross section
signals a change in the structure and/or
propagation of the $\Delta$ in medium beyond the smearing effect predicted by Fermi motion that 
is consistent with a downwards mass shift of the $\Delta$ and which 
needs to be further understood.
New measurements with heavier polarised targets, e.g., $^6$Li and $^7$Li, might 
be used to study the $A$ dependence of the effect connected to the M1 magnetic transition.
Also, it would be good to extend the energy range of the experiment 
to study the effect of heavier resonance excitations,
their mass shifts and widths as well as any change in the pion production threshold from individual bound nucleons.

\section*{Acknowledgments}

We thank 
H. Arenh\"ovel,
M. Bashkanov and V. Metag for helpful discussions.
We thank the ExtreMe Matter Institute EMMI at GSI, Darmstadt, for support in the framework of an EMMI Workshop EW22-03 {\it Meson and Hyperon Interactions with Nuclei}
during which this work has been initiated.
SDB thanks the Mainz Institute for Theoretical 
Physics (MITP) of the DFG Cluster of Excellence PRISMA*, 
Project ID 39083149, 
for its hospitality
where part of this work was completed.

\end{document}